# Signatures of Gate-Tunable Superconductivity in Trilayer Graphene/Boron Nitride Moiré Superlattice

Guorui Chen[1,2+], Aaron L. Sharpe[4,5+], Patrick Gallagher[1,2], Ilan T. Rosen[4,5], Eli Fox[4,9], Lili Jiang[2], Bosai Lyu[6,7], Hongyuan Li[6,7], Kenji Watanabe[8], Takashi Taniguchi[8], Jeil Jung[12], Zhiwen Shi[6,7], David Goldhaber-Gordon[5,9*], Yuanbo Zhang[3,7,10*], Feng Wang[1,2,11*]

[1]Materials Science Division, Lawrence Berkeley National Laboratory, Berkeley, CA, USA.
[2]Department of Physics, University of California at Berkeley, Berkeley, CA, USA.
[3]State Key Laboratory of Surface Physics and Department of Physics, Fudan University, Shanghai 200433, China.
[4]Department of Applied Physics, Stanford University, Stanford, CA 94305, USA.
[5]Stanford Institute for Materials and Energy Sciences, SLAC National Accelerator Laboratory, 2575 Sand Hill Road, Menlo Park, California 94025, USA.
[6]Key Laboratory of Artificial Structures and Quantum Control (Ministry of Education), School of Physics and Astronomy, Shanghai Jiao Tong University, Shanghai, China.
[7]Collaborative Innovation Center of Advanced Microstructures, Nanjing, China.
[8]National Institute for Materials Science, 1-1 Namiki, Tsukuba, 305-0044, Japan.
[9]Department of Physics, Stanford University, Stanford, California 94305, USA
[10]Institute for Nanoelectronic Devices and Quantum Computing, Fudan University, Shanghai 200433, China.
[11]Kavli Energy NanoSciences Institute at the University of California, Berkeley and the Lawrence Berkeley National Laboratory, Berkeley, CA, USA.
[12]Department of Physics, University of Seoul, Seoul 02504, Korea.

[+]These authors contributed equally to this work.
*Correspondence to: fengwang76@berkeley.edu, zhyb@fudan.edu.cn, goldhaber-gordon@stanford.edu

Understanding the mechanism of high temperature (high $T_c$) superconductivity is a central problem in condensed matter physics. It is often speculated that high $T_c$ superconductivity arises from a doped Mott insulator[1] as described by the Hubbard model[2–4]. An exact solution of the Hubbard model, however, is extremely challenging due to the strong electron-electron correlation. Therefore, it is highly desirable to experimentally study a model Hubbard system in which the unconventional superconductivity can be continuously tuned by varying the Hubbard parameters. Here we report signatures of tunable superconductivity in an ABC-trilayer graphene (TLG) / boron nitride (hBN) moiré superlattice. Unlike "magic angle" twisted bilayer graphene, theoretical calculations show that under a vertical displacement field the ABC-TLG/hBN heterostructure features an isolated flat valence miniband associated with a Hubbard model on a triangular superlattice[5,6]. Upon applying such a displacement field we find experimentally that the ABC-TLG/hBN superlattice displays Mott insulating states below 20 kelvin at 1/4 and 1/2 fillings, corresponding to 1 and 2 holes per unit cell, respectively. Upon further cooling, signatures of superconducting domes emerge below 1 kelvin for the electron- and hole-doped sides of the 1/4 filling Mott state. The electronic behavior in the TLG/hBN superlattice is expected to depend sensitively on the interplay between the electron-electron interaction and the miniband bandwidth, which can be tuned continuously with the displacement field D. By simply varying the D field, we demonstrate transitions from the candidate superconductor to Mott insulator and metallic phases. Our study shows that TLG/hBN heterostructures offer an attractive model system to explore rich correlated behavior emerging in the tunable triangular Hubbard model.

The ability to exfoliate and stack atomically thin two-dimensional (2D) materials into new classes of van der Waals heterostructures has ushered in a new era for synthetic quantum materials, where the properties of the materials can be conveniently controlled through both the composition and stacking order of different layered materials and an external electrical field from electrostatic gates. Such van der Waals heterostructures offer the possibility to design tunable material systems that exhibit fascinating new quantum phenomena. For example, correlated insulating states and unconventional superconductivity have recently been reported in magic-angle twisted bilayer graphene[7,8], and gate-tunable Mott insulating states have been observed in ABC trilayer graphene/hexagonal boron nitride (TLG/hBN) heterostructures[5]. In particular, the TLG/hBN system provides an ideal platform for systematic study of the triangular Hubbard model with 4-fold onsite degeneracy: theoretical calculations show that the system features an isolated nearly-flat miniband in a triangular superlattice, and the miniband's bandwidth can be tuned with a vertical electrical field. This is in contrast to the magic-angle twisted bilayer graphene system, where calculations show two nearly-flat minibands that always intersect in the single-particle bandstructure[8]. Here we report signatures of tunable superconductivity in a TLG/hBN heterostructure around the 1/4 filling Mott state, corresponding to one hole per unit cell in the moiré miniband. Two apparent superconducting domes are observed with electron and hole doping relative to the 1/4 filling Mott state, respectively. In addition, transitions between superconducting, insulating, and metallic states in the TLG/hBN heterostructure are readily controlled by a vertical electric field.

The sample fabrication process is similar to that reported in Ref. [8]. Briefly, near-field infrared (IR) nanoscopy is used to identify ABC and ABA regions in exfoliated trilayer graphene[9]. Dry transfer methods are used to pick up and assemble hBN/trilayer graphene/hBN stacks with careful angular alignment[10,11]. Standard e-beam lithography, reactive ion etching and electron-beam evaporation are used to fabricate TLG/hBN devices with a Hall bar geometry. The TLG is contacted through one-dimensional edge contacts with non-superconducting Cr/Au metals. We further deposit a metal top electrode to form dual-gated devices where the TLG/hBN heterostructures can be gated by both the top metal electrode and the bottom silicon substrate. Fig. 1a shows an optical image of a fabricated device. The dual-gate configuration allows us to control the carrier concentration and miniband bandwidth of the TLG/hBN heterostructure independently[12–14]: the vertical displacement field across the TLG is set by $D = \frac{1}{2}(D_b + D_t)$ and the charge concentration relative to the charge neutrality point by $n = (D_b - D_t)/e$, where $D_b = +\varepsilon_b(V_b - V_b^0)/d_t$ and $D_t = -\varepsilon_t(V_t - V_t^0)/d_t$ can be controlled by the bottom and top gate voltages, respectively. Here $\varepsilon_{b(t)}$ and $d_{b(t)}$ are the dielectric constant and thickness of the bottom (top) dielectric layers, and $V_b^0$ and $V_t^0$ are the effective offset in the bottom and top gate voltages caused by environment-induced carrier doping.

ABC-stacked TLG features a cubic energy dispersion and therefore a large effective mass at low energy[15–18]. In a TLG/hBN heterostructure in which the TLG is rotationally aligned to one of the hBN cladding layers, a moiré superlattice with a period of 15 nm folds the pristine TLG electronic band into a series of moiré minibands in the first moiré mini Brillouin zone[5,19–23]. Theoretical calculations show that the moiré superlattice in such a TLG/hBN heterostructure generates a periodic potential characterized by a triangular lattice, yielding first electron and hole minibands with very narrow energy bandwidth. (See SI section I and II for calculation details.) A schematic of the triangular TLG/hBN moiré superlattice is shown in Fig. 1b. Figure 1c displays the calculated energy dispersion of the few lowest electron and hole minibands for an ABC TLG/hBN heterostructure where the moiré superlattice is formed with the top hBN flake and the potential energy difference between the bottom and top graphene is $2\Delta = 20$ meV. This potential energy difference can be generated by a vertical displacement field of -0.4 V/nm[5], where the negative sign denotes the field pointing downward. In TLG/hBN system, the minibands around the K and K' valley in the original graphene Brillouin zone form time reversal pairs, where $E_K(p) = E_{K'}(-p)$. Due to the four-fold spin and valley-isospin degeneracy, 4 electrons are required to completely fill a miniband in TLG/hBN moiré superlattice.

Significant electron and hole asymmetry exists in the TLG/hBN system, and the first hole miniband tends to be narrower and better separated from the other bands. At $2\Delta = 20$ meV, the first hole miniband has a bandwidth $W$ as narrow as 15 meV, and is separated from other bands by over 10meV. The on-site Coulomb repulsion energy, on the other hand, can be estimated by $U \sim \frac{e^2}{4\pi\varepsilon_0 \varepsilon L_M}$. For $L_M$ = 15 nm and an hBN dielectric constant $\varepsilon = 4$, $U$ is around 25 meV, which is larger than the value of $W$. This dominant on-site Coulomb repulsion can lead to Mott insulator states in the flat and isolated hole miniband[5,24,25]. Experimentally, the gate-dependent four-probe resistance ($R_{xx}$) at a vertical displacement field of -0.4 V/nm in a TLG/hBN superlattice is shown in Fig. 1d for temperature 5 K. Prominent Mott insulating states are observed at 1/4 and 1/2 fillings of the hole miniband, corresponding to one hole and two holes per superlattice unit cell, respectively. The two-dimensional color plot of $R_{xx}$ as a function of $V_t$ and $V_b$ in Fig. 1e shows the evolution of the charge neutral point (CNP), 1/4 filling, 1/2 filling, and full filling (FFP) resistance peaks with the displacement field $D$. For relatively large $D$ field, resistivity peaks can be clearly identified at 1/4 and 1/2 filling of the first hole miniband. From both the experimental and calculated results, the electron and hole minibands are asymmetric with displacement field in the TLG/hBN heterostructure. As suggested above, the different behavior between the positive and negative displacement field arises from the fact that the moiré superlattice exists only at top hBN/TLG interface in this device.

Signatures of superconductivity emerge in this device below 1 kelvin. We first focus on a state close to the well-developed 1/4 filling Mott insulator with $D$ = -0.54 V/nm and $n$ = -5.4 × $10^{11}$ cm$^{-2}$. The first sign of superconductivity is a sharp drop of $R_{xx}$ from ~5 kΩ to ~300 Ω within the narrow temperature range of 2 K to 0.2 K. The lowest resistance then remains constant down to 0.04 K. An empirical fit to the Aslamazov-Larkin formula[26] of the $R_{xx}$ - $T$ curve, shown as the solid line in Fig. 2a, yields an estimated superconducting T$_c$ of 0.65 K. Non-zero residual resistance can appear in measurements of microscopic superconducting samples with poor electrical contacts, as reported for example in some magic-angle twisted bilayer graphene samples[27] and other 2D superconductors[28,29]. The residual resistance may also have a contribution from non-equilibrium quasi-particles in microscopic devices[30].

A second signature of superconductivity comes from measurements of the current-voltage relationship (*I-V* curves), as displayed in Fig. 2b. At the lowest temperatures, the *I-V* curves show a plateau below the critical current at ~ 10 nA. The plateau region tilts and exhibits nearly linear behavior at higher temperature. Figure 2c shows the differential resistance d*V*/d*I* as a function of driving current, which provides a better visualization of the critical supercurrent of ~ 10 nA below 0.3 K and the evolution to a normal metal behavior above ~ 1 K.

Figure 2d displays the critical supercurrent behavior in d*V*/d*I-I* curves as a function of the perpendicular magnetic field $B_\perp$ at a base temperature of 40 mK. There is a clear suppression of the apparent superconductivity by the magnetic field, and it almost disappears at $B_\perp$ ~ 0.7 T. Between 0.7 to 2 T, the differential resistance at low bias is relatively small and it exhibits a weak nonlinear *I-V* behavior (Fig. 2e). However, further measurement of *R-T* response at such magnetic fields shows a behavior characteristic of a metallic state with a weak temperature dependence (See SI section III).

Figure 2f further displays the in-plane magnetic field dependence of the critical supercurrent behavior at 40 mK. The superconductivity in TLG is again suppressed by the in-plane magnetic field $B_\parallel$, and disappears below 1 T. At the same time, we observe an anomalous resistance peak close to zero current bias at large in-plane magnetic field. This behavior is rather unusual. We do not know its origin and further experimental and theoretical studies will be needed to fully understand this in-plane magnetic field dependence.

Next we examine the superconductivity phase diagram as a function of temperature, *T,* and carrier doping, *n*. We fix *D* at two different values, *D* = -0.54 V/nm where both the 1/2 and 1/4 fillings Mott states appear (Fig. 3a) and *D* = -0.17 V/nm where only the 1/2 filling Mott state appears (Fig. 3b.)

At $D$ = -0.54 V/nm, two apparent superconducting domes emerge at low temperature near the 1/4 filling Mott state: superconductivity appears to exist for both electron and hole doping relative to the 1/4 filling Mott state, analogous to the behavior in high-$T_c$ cuprates[3]. This also resembles the behavior observed in magic angle twisted bilayer graphene around 1/2 hole filling[8], though a more recent work suggests that lower disorder produces only one dome, for hole doping relative to the Mott insulator[27]. The presence of superconductivity even right at 1/4 filling in our measurements can be attributed to the charge inhomogeneity in the trilayer graphene devices, which can lead to patches of superconductivity when the average filling corresponds to a Mott insulator as seen at higher temperatures (See SI section IV). The behavior around the 1/2 filling Mott state is more complex. The resistance remains rather high on the electron doping side of the 1/2 filling point. The $R$-$T$ curve for hole doping relative to the 1/2 filling Mott state is consistent with a superconducting transition, but the $I$-$V$ curve shows a very weak plateau. (See SI section V). Further studies will be required to conclusively establish the nature of this state.

At $D$ = -0.17 V/nm, where only the 1/2 filling Mott state exists, the phase diagram shows no superconductivity even at base temperature. The metal phases show very small resistance values at base temperature, but they can be distinguished from superconducting phases by their $R$-$T$ dependences and the lack of supercurrent behavior in the $I$-$V$ and d$V$/d$I$ curves.

The TLG/hBN system offers a platform to investigate the evolution of superconductivity where the bandstructure of the miniband can be continuously tuned by a vertical displacement field $D$. We fix the carrier concentration at a constant $n$ = -5.2 × $10^{11}$ cm$^{-2}$, which corresponds to a small electron doping relative to the 1/4 filling Mott states, and examine the electronic phases at different $D$. The four-probe resistance $R_{xx}$ as a function of $D$ and $T$ is displayed as a two-dimensional color plot in Fig. 4a. At small $D$ where the miniband bandwidth is relatively broad, the system exhibits a metallic phase. Figure 4b shows $R_{xx}$-$T$ plot at $D$ = 0, where the resistance is low and constant, suggesting that impurity scattering dominates at very low temperature. When $D$ is increased to a positive value, we observe a phase transition from the metallic state to a Mott insulating state due to the field-induced narrowing of the hole miniband[5]. A line cut at $D = 0.45$ V/nm of Fig. 4a shows the insulating $R_{xx}$-$T$ behavior of such a Mott insulating state (Fig. 4c). Superconductivity, however, never appears in this parameter space with a positive $D$. When $D$ is varied to negative values, we observe an evolution from the metallic phase ($D > -0.28$ V/nm) to a candidate superconducting phase ($D < -0.53$ V/nm). (We have limited $|D|$ to 0.6 V/nm to avoid possible damage to the gate dielectrics.) The transition region between $D = -0.28$ and $-0.53$ V/nm exhibits rather complex behavior, and we refer to it as a "resistive state" because the overall resistance is relatively high compared with the metallic region. Figure 4e shows several $R$-$T$ curves with different behaviors in this transition region.

The TLG/hBN superlattice provides a unique model system to study the triangular Hubbard model with 4-fold onsite degeneracy, associated with an isolated and electrically controllable nearly-flat 4-fold degenerate miniband. In this system, we experimentally find tunable Mott insulator states and signatures of tunable superconductivity. Further studies of such a tunable quantum system may shed light on to the longstanding question of high-$T_c$ superconductivity's relation to the Hubbard model. TLG/hBN systems also hold the promise to reveal completely new types of electronic states, such as spin liquid phases[25], electrically tunable Chern bands[31,32], and topological triplet superconductivity[24,33], all of which have been recently predicted for a triangular Hubbard model based on TLG/hBN superlattices.

**Supplementary Information**

Supplementary Information is available in the online version of the paper.

**Data Availability**

The data that support the findings of this study are available from the corresponding authors upon reasonable request.

**Methods**

**Sample fabrication.** TLG and hBN are mechanically exfoliated on $SiO_2$/Si substrate and the layer number of TLG is identified by optical contrast and AFM. ABC-TLG is characterized by near-field infrared nanoscopy and isolated in-situ by cutting with an AFM. hBN flakes are selected to be thicker than 30nm and without step edges. We create the hBN/TLG/hBN heterostructure by stacking different layers with a dry transfer method[26]. We identify the crystal orientation of TLG and hBN using the crystalline edges of the flakes, and manually align the TLG lattice with the hBN flake during the transfer process. The device is then etched into a Hall bar structure using standard e-beam lithography. The TLG is contacted through one-dimensional edge contacts with Cr/Au electrodes. We further deposit a metal top electrode to form a dual-gate device where the TLG/hBN heterostructure can be gated by both the top metal electrode and the bottom silicon substrate.

**Transport measurements.** The device is measured in a dilution refrigerator which achieves a base electron temperature of $T$ = 0.04 K, as determined by Coulomb blockade thermometry. Low temperature electronic filtering, including microwave filters, low-pass RC filters, and thermal meanders, are used to anchor the electron temperature as well as to prevent quasiparticle excitations from high frequency noise. Stanford Research Systems SR830 lock-in amplifiers with NF Corporation LI-75A voltage preamplifiers are used to measure the longitudinal resistance $R_{xx}$ of the device with an AC bias current of 0.5 nA to 1 nA at a frequency of 7 Hz. A Yokogawa 7651 DC voltage source is used in combination with a 100 MOhm bias resistor to add a DC bias current to the device. The voltage in the $I$-$V$ measurements is measured using an Agilent 34401A. In the d$V$/d$I$-$I$ measurements, a small AC current bias (0.5 nA) is generated by the lock-in amplifier output voltage in combination with a 1 GOhm bias resistor. This small AC current is added on top of the larger DC current bias, and the induced differential voltage is measured using the lock-in technique. In-plane magnetic field measurements are performed using an attocube atto3DR two-axis piezo rotator to control the sample orientation with respect to the field.

**Author contributions**
F.W., Y.Z. and D.G.-G. supervised the project. G.C. fabricated samples and performed basic transport characterizations at temperature above 5 K. G.C., A.S., P.G., I.R., and E.F. performed ultra-low temperature transport measurements. G.C., L.J., B.L., H.L. and Z.S. prepared trilayer graphene and performed near-field infrared and AFM measurements. K.W. and T.T. grew hBN single crystals. J.J. Calculated the band structures. G.C., A.S, P.G, I.R., D.G.-G., Y.Z. and F.W. analyzed the data. G.C. and F.W. wrote the paper, with input from all authors.

**Acknowledgments**
We acknowledge helpful discussions with Guangming Zhang and Tao Xiang. G.C. and F.W. were supported as part of the Center for Novel Pathways to Quantum Coherence in Materials, an Energy Frontier Research Center funded by the U.S. Department of Energy, Office of Science, Basic Energy Sciences. A.S. was supported by a National Science Foundation Graduate Research Fellowship and a Ford Foundation Predoctoral Fellowship. I.R., E.F., and D.G.-G.'s work on this project were supported by the U.S. Department of Energy, Office of Science, Basic Energy Sciences, Materials Sciences and Engineering Division, under Contract No. DE-AC02-76SF00515. Dilution fridge support: Low-temperature infrastructure and cryostat support were funded in part by the Gordon and Betty Moore Foundation through Grant No. GBMF3429. Part of the sample fabrication was conducted at Nano-fabrication Laboratory at Fudan University. Y.Z. acknowledges financial support from National Key Research Program of China (grant nos. 2016YFA0300703, 2018YFA0305600), NSF of China (grant nos. U1732274, 11527805, 11425415 and 11421404), and Strategic Priority Research

Program of Chinese Academy of Sciences (grant no. XDB30000000). Z.S. acknowledges support from National Key Research and Development Program of China (grant number 2016YFA0302001) and National Natural Science Foundation of China (grant number 11574204, 11774224). K.W. and T.T. acknowledge support from the Elemental Strategy Initiative conducted by the MEXT, Japan and the CREST (JPMJCR15F3), JST. J.J. was supported by the Samsung Science and Technology Foundation under Project No. SSTF-BA1802-06, and by the Korean National Research Foundation grant NRF-2016R1A2B4010105.

**Author Information**

The authors declare no competing financial interests. Correspondence and requests for materials should be addressed to F.W. (fengwang76@berkeley.edu), Y.Z. (zhyb@fudan.edu.cn), and D.G.-G. (goldhaber-gordon@stanford.edu).

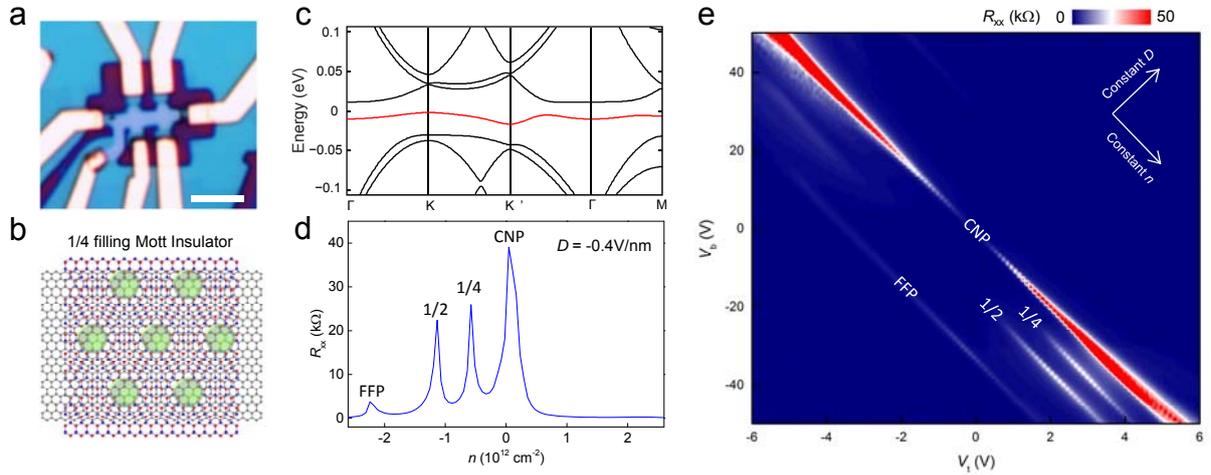

**Figure 1. Mott insulator in trilayer graphene/hBN moiré superlattice. a**, An optical image of the ABC TLG/hBN device with top and bottom gates. The white scale bar is 3 μm. **b**, A schematic of the triangular TLG/hBN moiré superlattice and the 1⁄4 filling Mott insulating state, which corresponds to one hole per superlattice unit cell. **c**, The single-particle energy dispersion of the lowest electron and hole minibands in the ABC TLG/hBN superlattice with an effective potential energy difference between the bottom and top graphene layer of $2\Delta = 20$ meV, which can be generated by a vertical displacement field of -0.4 V/nm. It features a narrow and isolated hole miniband, shown highlighted in red. **d**, $R_{xx}$ as a function of carrier density shows prominent Mott insulating states at 1/4 and 1/2 fillings with $D$ = -0.4 V/nm at $T$ = 5K. **e**, Two-dimensional color plot of $R_{xx}$ as a function of $V_t$ and $V_b$ at $T$ = 5 K. The resistance peaks at 1/4 and 1/2 fillings of the first hole miniband can be clearly identified for relatively large $D$ field.

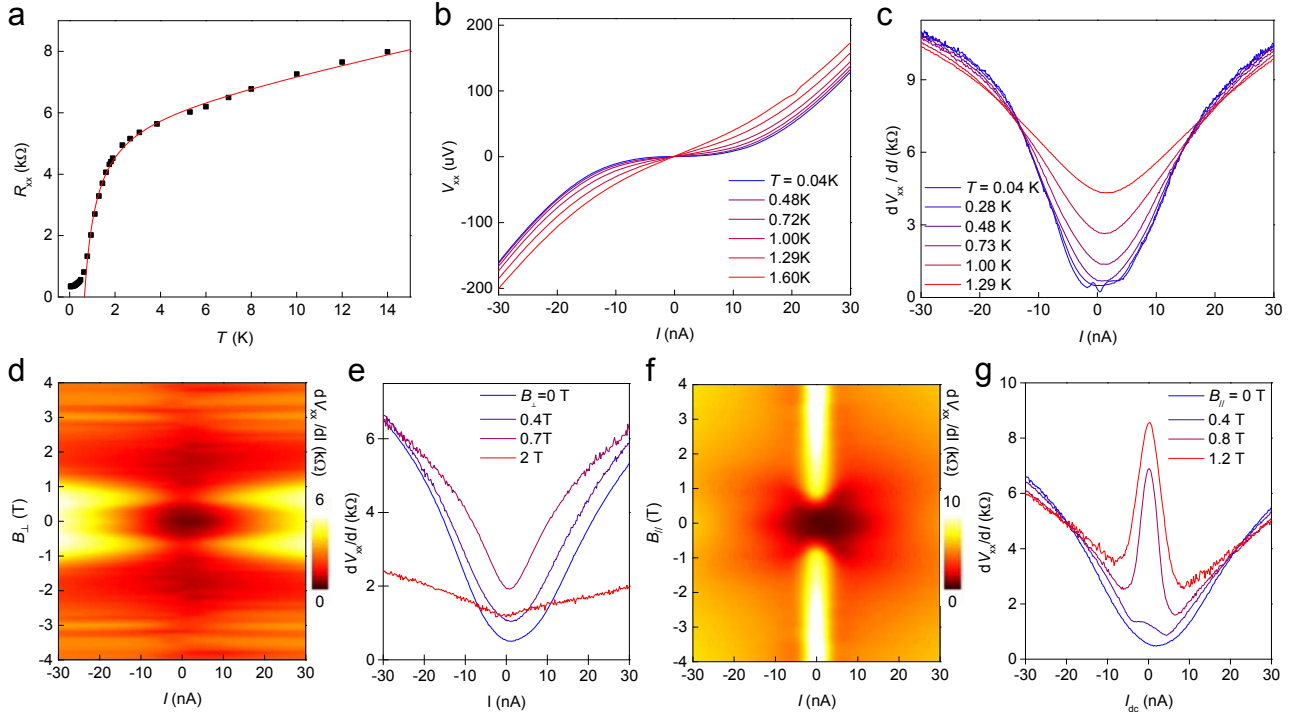

**Figure 2. Superconductivity in TLG/hBN. a**, $R_{xx}$-$T$ curve at $D$ = -0.54 V/nm and $n$ = -5.4 × $10^{11}$ cm$^{-2}$ shows characteristic behavior of a superconducting transition. An empirical fit to the Aslamazov-Larkin formula (red line) yields an estimated superconducting temperature of 0.65 K. **b**, *I-V* curves at different temperatures show a plateau below the critical current at ~ 10 nA for temperatures below 0.3 K. This plateau region tilts and becomes close to linear at higher temperature, characteristic of a superconducting transition. **c**, d*V*/d*I-I* curves at different temperatures. A critical current of ~ 10 nA is observed at the lowest temperatures. **d**, The d*V*/d*I* color plot as a function of dc bias current and perpendicular magnetic field at *T* = 0.04 K. **e**, Line cuts of **d** at $B_\perp$ = 0, 0.4, 0.7 and 2 T. **f**, The d*V*/d*I* color plot as a function of dc bias current and in-plane magnetic field at *T* = 0.04 K. **g**, Line cuts of **d** at $B_{//}$ = 0, 0.4, 0.8 and 1.2 T. Symmetrized $R_{xx}$ (B) = [$R_{xx}$ (+B) + $R_{xx}$ (-B)]/2 is presented to get rid of possible $R_{xy}$ component in **d-g**.

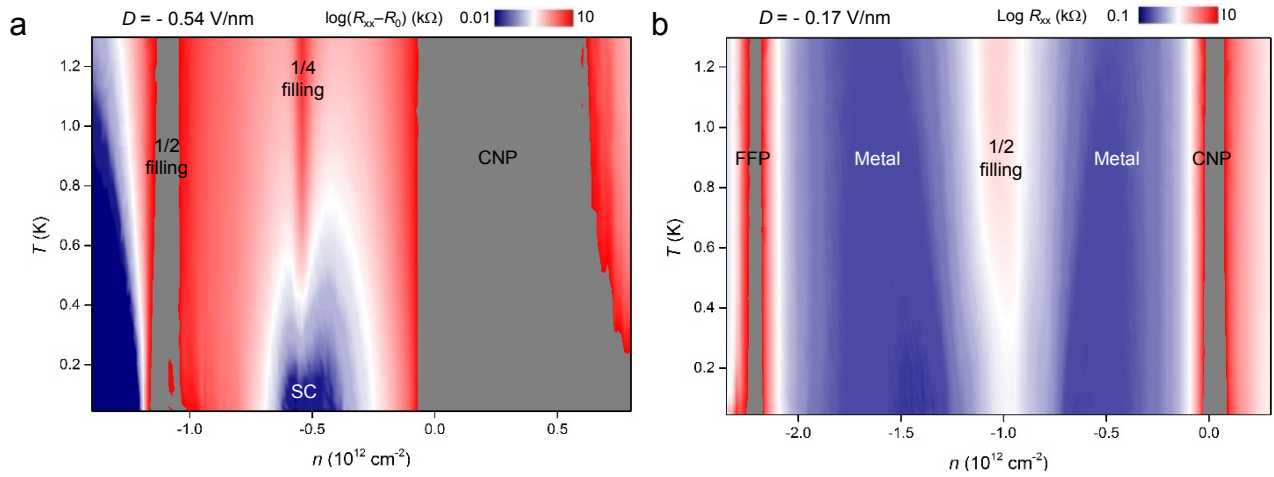

**Figure 3. Carrier density dependent phase diagram.** $R_{xx}$ as a function of carrier density and temperature at $D = -0.54$ V/nm (**a**) and $D = -0.17$ V/nm (**b**). Superconductor phase emerges at low temperature near the 1⁄4 Mott state for $D = -0.54$ V/nm. Only the 1/2 filling Mott state exists for $D = -0.17$ V/nm, and no superconductivity state exists even at base temperature. Both the superconducting phase and metal phase show very small resistance values at base temperature, but they can be distinguished by the supercurrent behavior in the *I-V* and d*V*/d*I* curves and by *R-T* dependence (Fig. 4b vs Fig. 4d). Color scale: $R_0 = 380$ Ω.

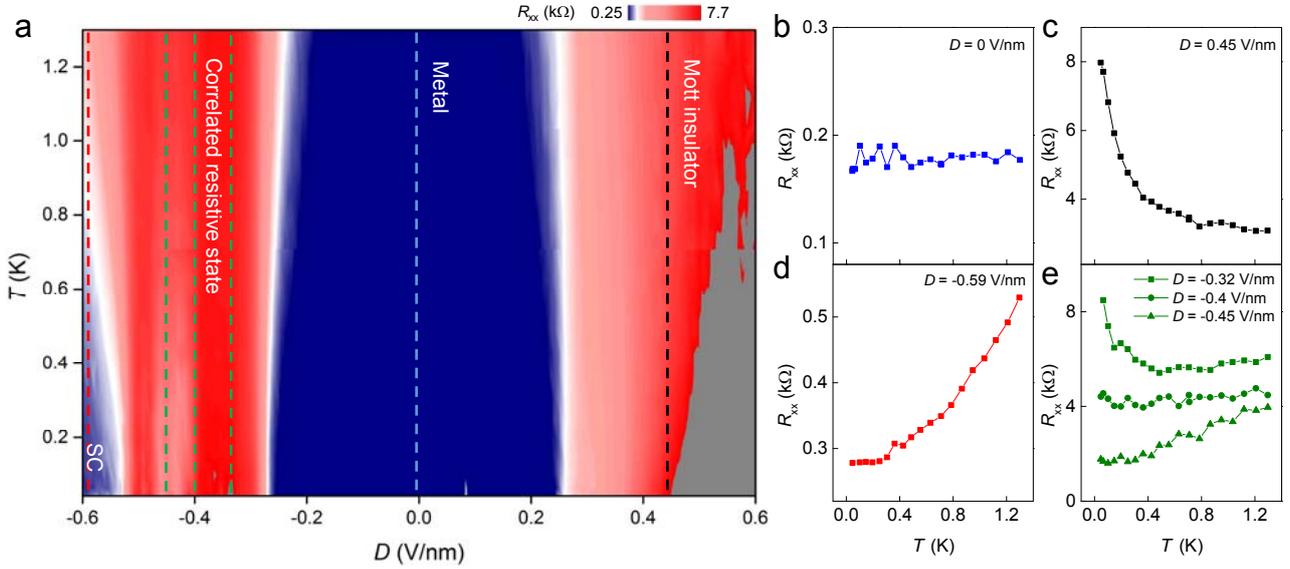

**Figure 4. Tunable electronic phases with the displacement field. a**, $R_{xx}$ as a function of $D$ and $T$ at fixed doping $n = -5.2 \times 10^{12}$ cm$^{-2}$ relative to CNP, corresponding to slight electron doping relative to 1/4 hole filling. $D$ modifies the bandstructure of the minibands in TLG/hBN, therefore the charge correlations. As a function of $D$, the system can be tuned across four different electronic states from left to right: superconductivity (SC), correlated resistive state, metal, and Mott insulator. **b-e**, Vertical line cuts of **a** at selected $D$ values to illustrate the $R_{xx}$-$T$ behavior of different electronic states. **b**, The metallic state at $D = 0$ V/nm shows a low and constant resistance due to the dominating impurity scattering. **c**, The Mott insulator state at $D = 0.45$ V/nm shows an increased resistance at lower temperature. **d**, The superconducting state at $D = -0.59$ V/nm shows a rapidly decreasing resistance to a constant residue value at low temperature. **e**, The transition region between $D = -0.28$ to $-0.53$ V/nm exhibits rather complex behavior, and we refer it as a "correlated resistive state" because the overall resistance is relatively high compared with the metallic region. The state at $D = -0.32$ V/nm shows a weak insulator-like behavior with increases resistance at the lowest temperatures, the state at $D = -0.4$V/nm shows an almost constant but relatively large resistance value, while the state at $D = -0.45$ V/nm displays a strange-metal-like behavior with an almost linear decrease of resistance at low temperature.